\documentclass[letterpaper,10pt,twocolumn]{article} %%LaTeX 2e
\usepackage{amsmath,amsthm,amssymb,amsfonts}
\usepackage{graphicx,float,mathcomp}
\usepackage{subfigure}

\usepackage[top= 0.75in, bottom=1in, left= 0.625in, right= 0.625in]{geometry}

{
}

\usepackage{amsmath}
\usepackage{amssymb}
\usepackage{graphicx}
\usepackage{graphicx}% Include figure files
\usepackage{dcolumn}% Align table columns on decimal point
\usepackage{bm}% bold math
\usepackage{amsfonts}
\usepackage{latexsym}
\usepackage{pdfsync}
    \usepackage{fullpage} % For More Realistic Page Usage
\usepackage{colordvi}
\usepackage{color}
\usepackage{booktabs}
\usepackage{graphicx}
\usepackage{subfigure}
\input xy
\xyoption{all}

\begin{document}

\newcommand{\commentout}[1]{}
\newcommand{\nwc}{\newcommand}
\nwc{\nn}{\nonumber}

\nwc{\nwt}{\newtheorem}
\nwt{cor}{Corollary}
\nwt{proposition}{Proposition}
\nwt{corollary}{Corollary}
\nwt{theorem}{Theorem}
\nwt{summary}{Summary}
\nwt{lemma}{Lemma}
\nwt{definition}{Defintion}
\nwt{remark}{Remark}

\nwc{\CC}{\mathbb{C}}
\nwc{\ZZ}{\mathbb{Z}}
\nwc{\RR}{\mathbb{R}}
\nwc{\ep}{\epsilon}

\nwc{\beq}{\begin{eqnarray}}
\nwc{\eeq}{\end{eqnarray}}

\pagestyle{empty}

\def\refname{\centerline{\footnotesize  \rm REFERENCES}}

\title{\Huge Super-Resolution by Compressive Sensing Algorithms}

\author{ Albert  Fannjiang\thanks{\footnotesize Corresponding author:
fannjiang@math.ucdavis.edu. The research is partially supported by
the NSF under grant DMS - 0908535.}, \ Wenjing Liao\\
{\normalsize Department of Mathematics}\\
{\normalsize UC Davis, CA 95616-8633.}}
\date{}
\maketitle
\thispagestyle{empty}

\pagestyle{empty}

{\bf \small
{\em Abstract -} 
In this work, 
super-resolution by 4 compressive sensing methods (OMP, BP, BLOOMP, BP-BLOT)  with highly coherent {\em partial}  Fourier measurements
is comparatively  studied. 

An alternative metric more suitable for gauging the quality of spike recovery
is introduced and based on the concept of filtration with a parameter representing 
the level of tolerance for  support offset. 

In terms of the filtered error norm
only BLOOMP and BP-BLOT can perform
grid-independent recovery of  well separated spikes of Rayleigh index $1$ for arbitrarily
large super-resolution factor. 
Moreover both BLOOMP and BP-BLOT can localize
spike support within a few percent of the Rayleigh length.
This is a weak form of super-resolution. 

Only BP-BLOT can achieve
this feat for closely spaced spikes separated by a fraction of the Rayleigh length,  a strong form of super-resolution. 
%The quality of spike value recovery can be further improved by increasing the number
%of Fourier data within the same bandwidth, thus moving from the under-sampling 
 % to the full and over-sampling regimes.   
}\\

\centerline{\footnotesize I. INTRODUCTION}

Superresolution  as Fourier spectrum extrapolation (i.e.  uncovering high spatial frequency components from low spatial frequency data) is  typically an ill-posed process and prone to extreme instability to  noise, unless additional prior knowledge and/or multiple data sets are available. 
Many techniques have been proposed in the literature but few have robust performances
and rigorous  foundation. 

%Compressive sensing (CS) algorithms are often believed to possess super-resolution capability. There are, however, several aspects of super-resolution.
A basic result in super-resolution with Fourier measurements is given by Ref. \cite{Donoho}  (see also Ref. \cite{Candes}). 
In the present work we discuss the strengths and weaknesses  of this result and give a numerical assessment of 
the super-resolution capability of compressive sensing (CS) algorithms.
 
First let us review the set-up of Fourier measurements of grid-bound spikes.
Consider the noisy Fourier data $y=\Phi x+e$ 
of a { grid-bound} spike train  
\begin{equation}
\label{spike}
x(t) = \sum_{l=0}^{N-1} x(l)\delta(t-\frac{l}{N})
\end{equation}
 on a fine grid of spacing $N^{-1}$ with the sensing matrix element
 \beq
 %y(k) & = \int_0^1 e^{- 2\pi i k t} x(t) dt + e(k) \\
\Phi_{kl} & = e^{-2\pi i k l / N}, \quad  k \in [0, N/F).\label{width}
\eeq
%where $k$ needs not be an integer. 

%Let $\mathbf{x} = \{x(l)\}\in \CC^N, y= \{y(k)\} \in \CC^n$ and $e= \{e(k)\} \in \CC^n$. $A \in \CC^{n \times N}$ is defined to be the sensing matrix with $A_{kl} = \text{exp}(- 2 \pi k l/ N)$, and then $y = A x + e$. 
 When  $F>1$, one is confronted with the problem of retrieving the fine scale structure of $x$ from a coarse scale information only.  The Rayleigh resolution length  $\ell$ associated with the Fourier data is the reciprocal of the observed  bandwidth in (\ref{width}). The ratio $F$ of the Rayleigh length to the fine grid spacing
 is   the super-resolution factor Ref. \cite{Candes,Donoho} or the refinement factor Ref. \cite{SPIE11,blo}.

According to  Ref. \cite{Donoho}, 
 any grid-bound spike train of the form (\ref{spike})  that is  consistent with the Fourier  data over 
 the frequency band $[0, N/F)$ deviates from the true solution by
\begin{equation}
\label{error}
\hbox{Error $\leq$ Constant $\cdot F^\alpha \cdot$ Noise},\quad \alpha \leq 2R+1
\end{equation} where \beq
\label{ray}
R(S)=\min \left\{r: r\geq \sup_t \#(S\cap [t,t+4\ell r))\right\} 
\eeq is  the Rayleigh index of the support set $S$ of $x$.
As explained in Ref. \cite{Donoho}, a set has Rayleigh index at most $R$ if in any interval of
length 4$\ell\cdot R$ there are at most $R$ spikes.

To increase $R$, one can  decrease the minimum distance and/or increase the number of spikes separated below $4\ell$ on {\em (local)  average}. The power $\alpha$ measures the degree of noise amplification 
by super-resolution.  

Recently Ref. \cite{Candes},  an error bound like (\ref{error}) with $\alpha=2$  is established for
reconstruction of  spikes  separated by 
at least $4\ell$ (thus $R=1$) by convex programming analogous to  Basis Pursuit (BP) Ref. \cite{CDS}
with Fourier data sampled at integers $k$ of the band $[0,N/F)$.
%Moreover, this can be achieved with a sparse measurement. 

The above sensing set-up, however,  has the following two drawbacks. \\
%The error bound (\ref{error}) with $F>1$ but $R=1$ for BP indicates a {\em weak} form of  super-resolution. 

\centerline{\footnotesize II. GRID-INDEPENDENT CS}

The first drawback is that when the prior information
of {\em grid-bound} spikes is not available, then imposing such a model assumption can lead to  poor reconstruction. 

To reduce the gridding  error due to off-grid objects one can reduce the grid spacing. But to resolve the finer grids, one must increase the received bandwidth which in turn brings the gridding error back to the original level and defeats the purpose of grid refinement. The same problem arises in radar imaging when the probe wavelengths are 
larger than the sizes of the off-grid scatterers. Often brushed off in literature Ref. \cite{Seg09, Seg12}, this issue is particularly pertinent   to sparsity-based  imaging schemes relying  on {\em accurate} and {\em sparse} representation of the objects which, in view of the intrinsically discrete nature of  CS framework, may be utterly untenable 
for  imaging  in continuum.   

How to get out this conundrum? Recently we have developed a solution method based on the techniques of coherence Band exclusion and Local Optimization (acronym: BLO) Ref. \cite{blo} which
introduces a more refined notion of local coherence bands.  

Roughly speaking the idea is this: Let the standard grid spacing be the Rayleigh resolution length $\ell$ and the refined grid spacing be $\ell'<\ell$.   When the super-resolution factor $F=\ell/\ell'$ is large, the nearby columns are nearly parallel, forming the coherence bands. The size of
the coherence band is approximately twice the Rayleigh length.   Due to the presence of coherence bands, the mutual coherence of the sensing  matrix is close to one, leading to
a poor condition number of even small column sub-matrices. As expected standard CS methods break down in this
situation.

 If, however, the spikes are separated by at least  $\ell$ then spikes do not fall into the coherence band of one another. 
So if one searches  the next object  {\em outside} the coherence bands of the previously identified objects by a greedy algorithm, say Orthogonal Matching Pursuit (OMP),  then the coherence bands would not get in the way of reconstruction. This is the technique of coherence band exclusion.

To improve the accuracy of reconstruction, one can further zoom in  within coherence bands
of the recovered objects one at a time  by locally minimizing  the residual which is inexpensive computationally. This is the technique of local optimization  (see Algorithm 1). 
\begin{center}
   \begin{tabular}[width = 3in]{|l|}
   \hline  
  {{\bf Algorithm 1.}\quad  Local Optimization (LO)}  \\ \hline
    Input:$\Phi,y,  S^0=\{i_1,\ldots,i_k\}$.\\
Iteration:  For $n=1,2,...,k$.\\
1) $x^n= \hbox{arg}\,\,\min_{z}\|\Phi z-y\|,$\quad\\
\quad $ \hbox{supp}(z)=(S^{n-1} \backslash \{i_n\})\cup \{j_n\}, $\\
  \quad $  j_n\in \hbox{Band}(\{i_n\})$.
\\
 2) $S^n=\hbox{supp}(x^n)$.\\
    Output:  $S^k$.\\
    \hline
   \end{tabular}
\end{center}
\begin{remark}
For spikes separated by at least $\ell$, the ``Band($\{j_n\}$)" in Algorithm 1 is
the set of  indices within the distance $\ell$ from $\{j_n\}$. For spikes
separated by less than $\ell$, ``Band($\{j_n\}$)" is the set of
indices within half the minimum separation of spikes. 
In the latter situation, the algorithm requires the prior knowledge
of the minimum separation of spikes. 

The same remark applies to Algorithms 2 and 3 below. 

\end{remark}

In Ref. \cite{blo} we have obtained performance guarantee for the BLO-based greedy algorithm, BLOOMP (standing for BLO-based OMP), which can recover the support of spikes separated by at least $3\ell$ {\em within the accuracy of one $\ell$} (as measured by the Bottleneck distance)
{\em independent of the super-resolution factor}. In reality, the accuracy of the support estimate is  just a few percent of the Rayleigh length $\ell$ (see below). Note that spikes separated by at least $3\ell$ can
comprise a set of arbitrarily large Rayleigh index. 

Moreover, this grid-independent 
performance can be achieved with sparse measurements, leading to
grid-independent CS. The numerical performances of BLOOMP and other variants have been thoroughly and systematically
tested in Ref. \cite{blo}.

\begin{center}
   \begin{tabular}[width = 3in]{|l|}
   \hline
   
   {{\bf Algorithm 2.} BLOOMP} \\ \hline
   Input: $\Phi, y, s=\hbox{\rm sparsity (number of spikes)}$\\
 Initialization:  $x^0 = 0, r^0 = y$ and $S^0=\emptyset$ \\ 
Iteration: For  $n=1,...,s$\\
\quad {1)  $i_{\rm max} = \hbox{arg}\max_{i}|\Phi_i^*r^{n-1} | , i \notin \hbox{\rm Band}(S^{n-1}) $} \\
  \quad      2) $S^{n} = \hbox{LO}(S^{n-1} \cup \{i_{\rm max}\})$\\
   where $\hbox{LO}$ is the output of Algorithm 2.\\
  \quad  3) $x^n = \hbox{arg}  \min_z \|
     \Phi z - y\|$ s.t. \hbox{supp}($z$) $\in S^n$ \\
  \quad   4) $r^n = y- \Phi x^n$\\
 %5)  Stop if $\|\br^n\|_2\leq\ep$.\\
 Output: $x^s$. \\
 \hline
   \end{tabular}%\label{tab2}
\end{center}

When the BLO technique is combined with thresholding  we have the Band-excluded, Locally Optimized Thresholding (BLOT) which can be used to enhance the performance of BP. The BLOT-enhanced BP is called BP-BLOT (see Algorithm 3). 

\begin{center}
   \begin{tabular}[width=3in]{|l|}
   \hline
   
   {{\bf Algorithm 3.}\quad BLOT}  \\ \hline
    Input: $x$, $\Phi, y, s=\hbox{\rm sparsity (number of spikes)}$.\\
    Initialization:  $S^0=\emptyset$.\\
Iteration:  For $n=1,2,...,s$.\\
\quad 1) $i_n= \hbox{arg}\,\,\max_j |x_j|, j\not\in \hbox{Band}(S^{n-1}) $.\\
 \quad 2)  $S^n=S^{n-1}\cup\{i_n\}$.\\
    Output:\\
     $\hat x=\hbox{arg}\min \|\Phi z-y\|_2$, $\hbox{supp}(z)\subseteq \hbox{LO}(S^s),$\\
  where $\hbox{LO}$ is the output of Algorithm 1.\\
     \hline
    %$\hat\mbx= \mbx_{S^s}$ where $\mbx_{S^s}$ is the
    %Hadamard 
    %product of $\mbx$ with the indicator function of $S^s$.\\ .\\
   \end{tabular}
\end{center}

Note that  BLOOMP and BP-BLOT require the prior information of the sparsity. \\

%This does not mean, however,  that the BLO technique always improves the error bound in
%(\ref{error}), which  leads us to the next critique.  \\

\centerline{\footnotesize III. FILTERED ERROR METRIC }

The second drawback of (\ref{error}) is that  the {\em discrete} norm used in (\ref{error}) 
does not take into account of the {\em degree of separation}  between the estimated support and
the true support as  measured by the Hausdorff distance or the  Bottleneck distance.
The discrete norm treats any amount of support offset equally. 

An easy remedy to the injudicious treatment of support offset is to use instead 
 the {\em filtered error norm}  
$
\| \hat{x}_\eta - x_\eta\|,
$
where $x_\eta$ and $\hat{x}_\eta$ are, respectively,  $x$ and
$\hat{x}$ convoluted with an approximate delta-function of
width $2\eta$.

If every spike of $\hat{x}$ is within $\eta$ distance 
from a spike of $x$ {\em and}  if the amplitude differences are  small,
then  the $\eta$-filtered error is small. The filter parameter $\eta$  represents  the level of tolerance for
the support off-set in a specific context. The filtered error plot,
as $\eta$ and noise level vary, will give a  more accurate and complete
picture of the super-resolution effect.  We will demonstrate the utility 
of the filtered norm in the numerical tests. \\

\commentout{
\centerline{\footnotesize IV. WEAK VS. STRONG SUPERRESOLUTION}

The grid-independent performance guarantee in  Ref. \cite{blo} can only be a weak form
of super-resolution since separation by more than the Rayleigh length is assumed.

To test the strong form of super-resolution, we need to go for sub-Rayleigh separation
of spikes.  In this case  the coherence bands of spikes overlap. Hence the band-exclusion technique
in Algorithms 1,  2 and 3  need to be modified to exclude instead smaller zones 
of previously detected spikes. The size of the exclusion zones is determined by 
the prior knowledge of the least separation of the spikes. \\
}

\centerline{\footnotesize IV. NUMERICAL TESTS}

\commentout{

 \begin{figure}[t]
  \centering
  \subfigure[OMP]{
    \includegraphics[ width = 1.5in]{Demo/OMPF50Noise5.eps}}
      %%%%%%%%%%%%%%%%%%%%%%%%%%%%%%%%%
          \subfigure[BLOOMP]{
             \includegraphics[width = 1.5in ]{Demo/BLOOMPF50Noise5.eps}}
        %%%%%%%%%%%%%%%%%%%%%%%%%%%%%%%%%
          \subfigure[BP]{
    \includegraphics[width = 1.5in]{Demo/BPDNF50Noise5.eps}}
      %%%%%%%%%%%%%%%%%%%%%%%%%%%%%%%%%
          \subfigure[BP-BLOT]{
             \includegraphics[ width = 1.5in]{Demo/BPDNBLOTF50Noise5.eps}}
        %%%%%%%%%%%%%%%%%%%%%%%%%%%%%%%%%
          \caption{Reconstructions  of the real part of 20 spikes with $R=1$ (minimum distance $3\rho$), $F=50, \hbox{\rm SNR} = 20$. }
          \commentout{
          (a) OMP, residual $\approx 7\%$, unfiltered error $\approx 97\%$, $0.1$-filtered error $\approx 28\%$,
           (b) BLOOMP, residual $\approx 4\%$, unfiltered error $\approx 41\%$, $0.1$-filtered error $\approx 6\%$,
            (c) BPDN, residual $\approx 5\%$, unfiltered error $\approx 153\%$, $0.1$-filtered error $\approx 26\%$,
             (d) BPDN-BLOT, residual $\approx 4\%$, unfiltered error $\approx  41\%$, $0.1$-filtered error $\approx 6\%$.  }    
        \label{fig:R1}%% label for entire figure
\end{figure}
}

 \begin{figure}[t]
  \centering
  \subfigure[OMP]{
    \includegraphics[ width = 1.4in]{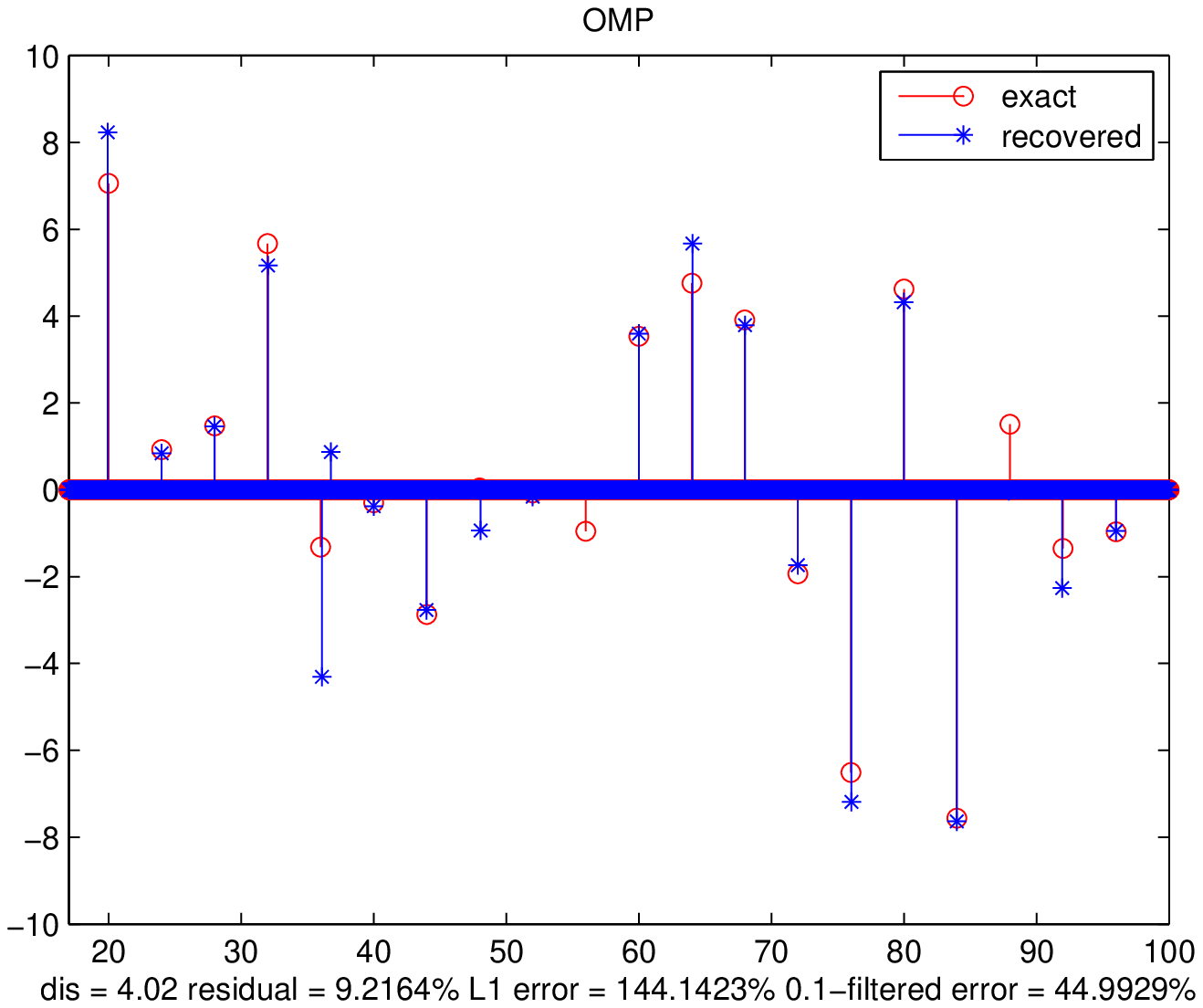}}
      %%%%%%%%%%%%%%%%%%%%%%%%%%%%%%%%%
          \subfigure[BLOOMP]{
             \includegraphics[width = 1.4in ]{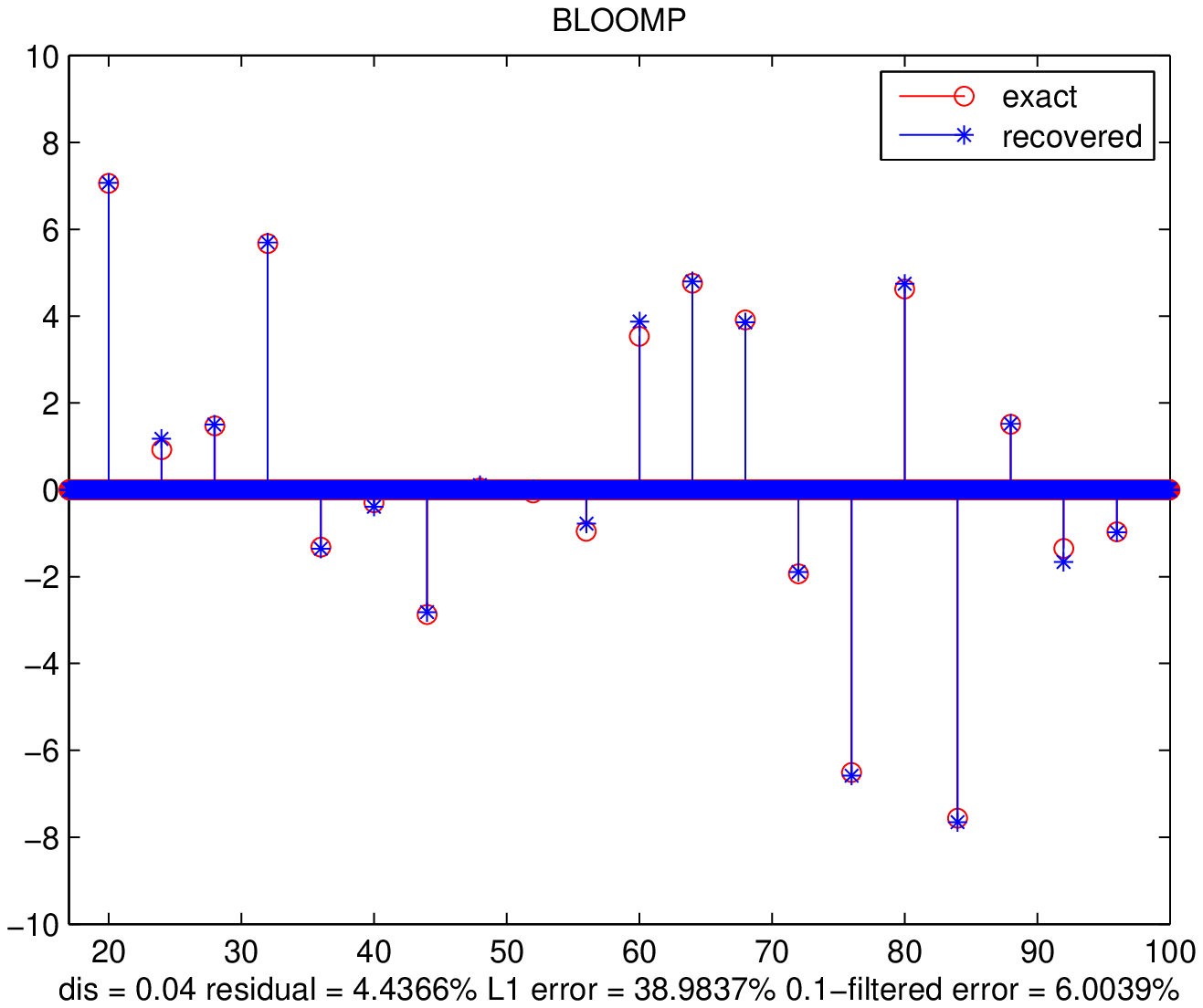}}
        %%%%%%%%%%%%%%%%%%%%%%%%%%%%%%%%%
          \subfigure[BP]{
    \includegraphics[width = 1.4in]{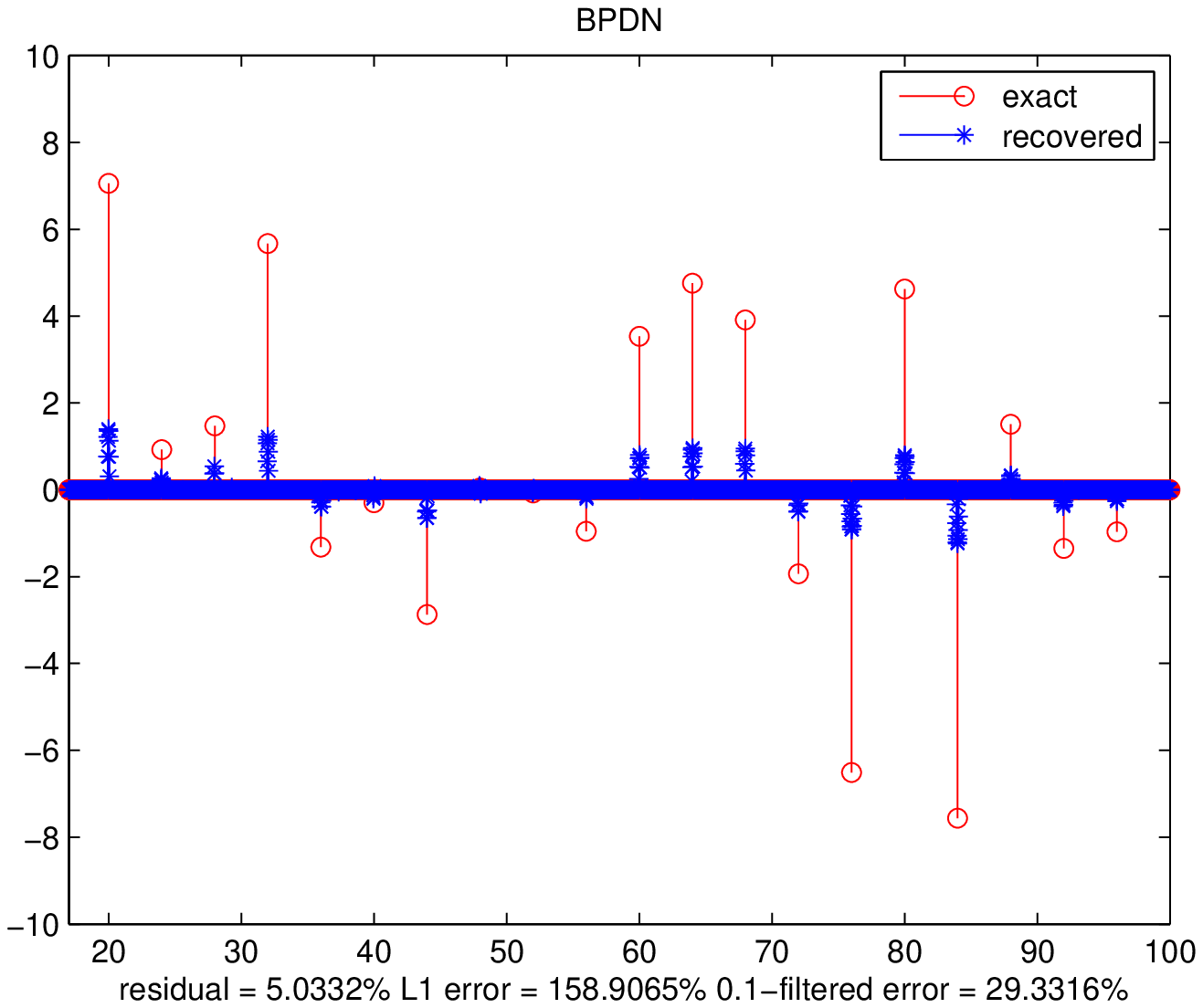}}
      %%%%%%%%%%%%%%%%%%%%%%%%%%%%%%%%%
          \subfigure[BP-BLOT]{
             \includegraphics[ width = 1.4in]{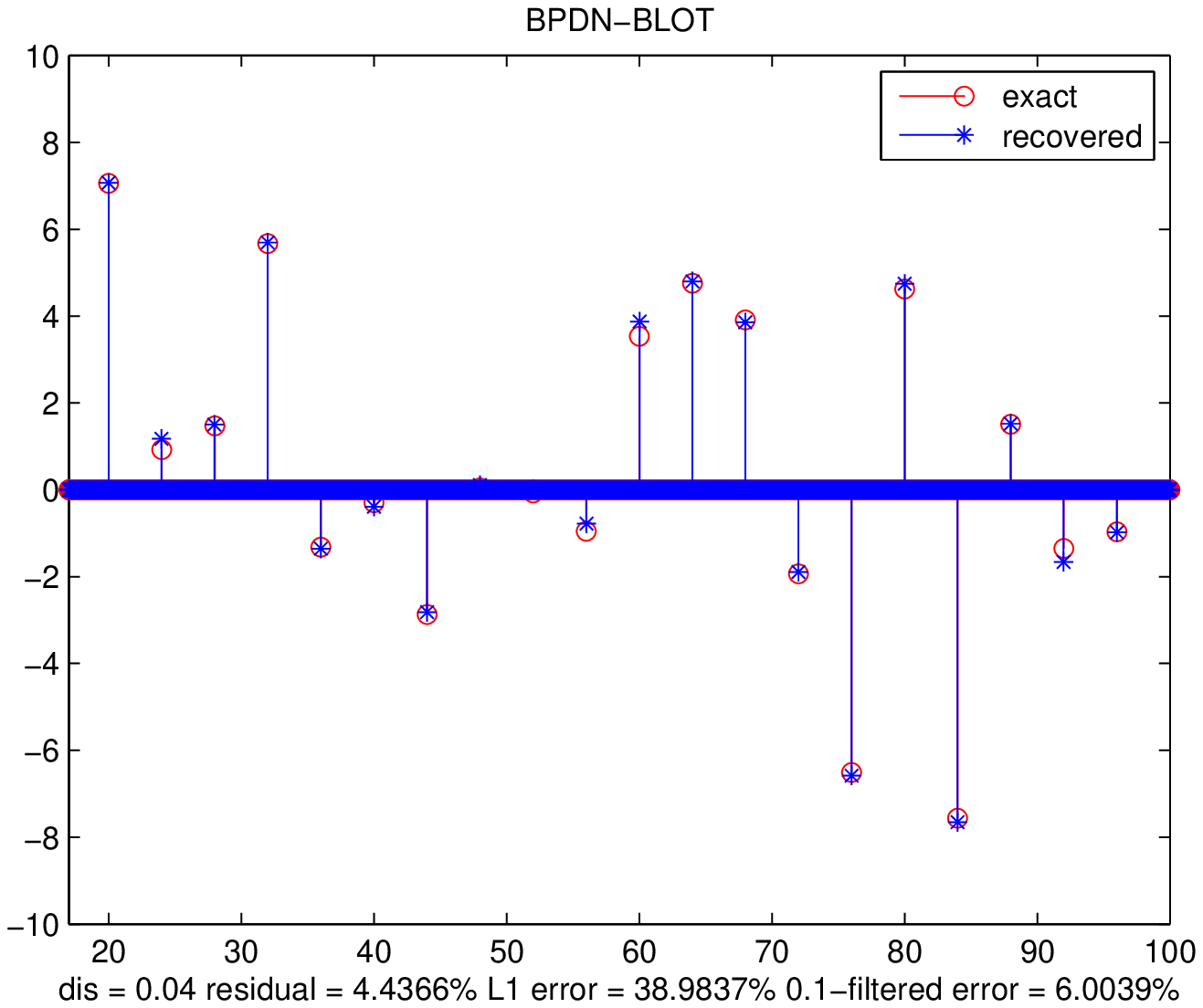}}
        %%%%%%%%%%%%%%%%%%%%%%%%%%%%%%%%%
          \caption{Reconstructions  of the real part of 20 randomly phased spikes with $R=1$ (minimum distance $4\ell$), $F=50, \hbox{\rm SNR} = 20$. }
          \commentout{
          (a) OMP, residual $\approx 7\%$, unfiltered error $\approx 97\%$, $0.1$-filtered error $\approx 28\%$,
           (b) BLOOMP, residual $\approx 4\%$, unfiltered error $\approx 41\%$, $0.1$-filtered error $\approx 6\%$,
            (c) BPDN, residual $\approx 5\%$, unfiltered error $\approx 153\%$, $0.1$-filtered error $\approx 26\%$,
             (d) BPDN-BLOT, residual $\approx 4\%$, unfiltered error $\approx  41\%$, $0.1$-filtered error $\approx 6\%$.  }    
        \label{fig:R1}%% label for entire figure
\end{figure}

In all our tests, we use $150\times N$  partial Fourier matrices (\ref{width}) where
$N=150F $ for various $F$
and  $k=0,1,2,\cdots, 149.$ In this setting, $\ell=1/150$. 

 For a  demonstration of grid-independent recovery, Fig.\ref{fig:R1} shows reconstructions of
20 spikes separated by at least $4\ell$ ( $R=1$) by using OMP, BP, BLOOMP and BP-BLOT with noisy ($5\%$) Fourier  data. For this simulation, $F=50$. We use the open-source code YALL1 ({\tt http://yall1.blogs.rice.edu/})  to find the BP solution. 

In this test, OMP tends to  miss small spikes. BLOOMP, however, approximately recover
the support and magnitudes of the spikes. While the  BP
reconstruction tends to cluster around the true spikes, BP-BLOT dramatically improves the performance. 
BLOOMP and BP-BLOT have a similarly superior
performance which is essentially independent of $F$. BLOOMP and BP-BLOT also perform
much better than other existing schemes (see Ref. \cite{blo, SPIE11} for systematic comparison). 

More quantitatively, the BLO technique reduces the unfiltered error
144\%  and $0.1$-filtered error 45\% for OMP to 39\% and 6\%, respectively, for BLOOMP.
The BLOT technique reduces the unfiltered error
159\% and $0.1$-filtered error 29\% for BP to
39\% and 6\%, respectively, for BP-BLOT. 

\begin{figure}[t]
  \centering
               %%%%%%%%%%%%%%%%%%%%%%%%%%%%%%%%%
             \subfigure[SNR=100, $\eta=0$]{
             \includegraphics[width = 1.4in]{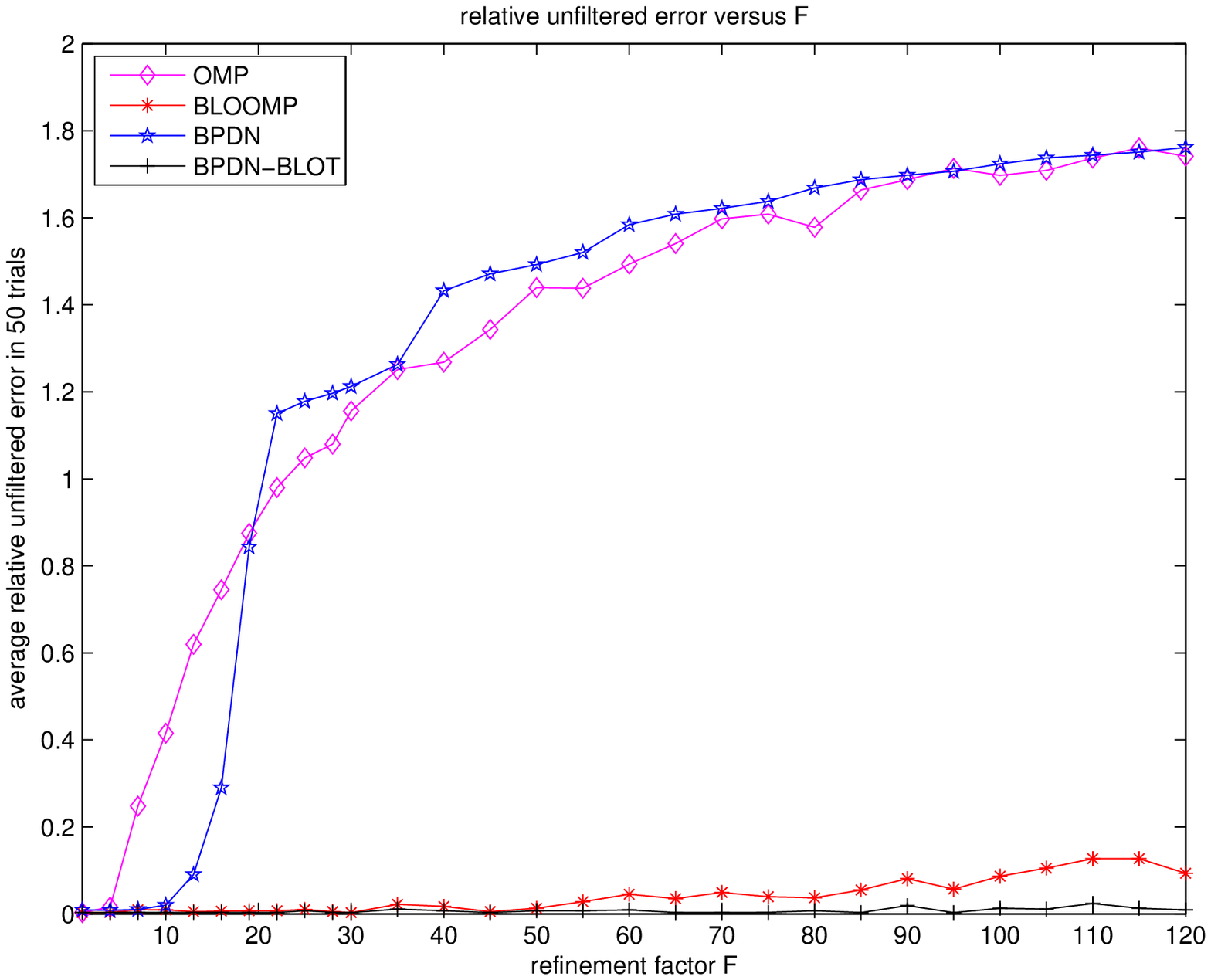}}
        %%%%%%%%%%%%%%%%%%%%%%%%%%%%%%%%%        
          \subfigure[SNR=100, $\eta=0.05\ell$]{
             \includegraphics[width = 1.4in]{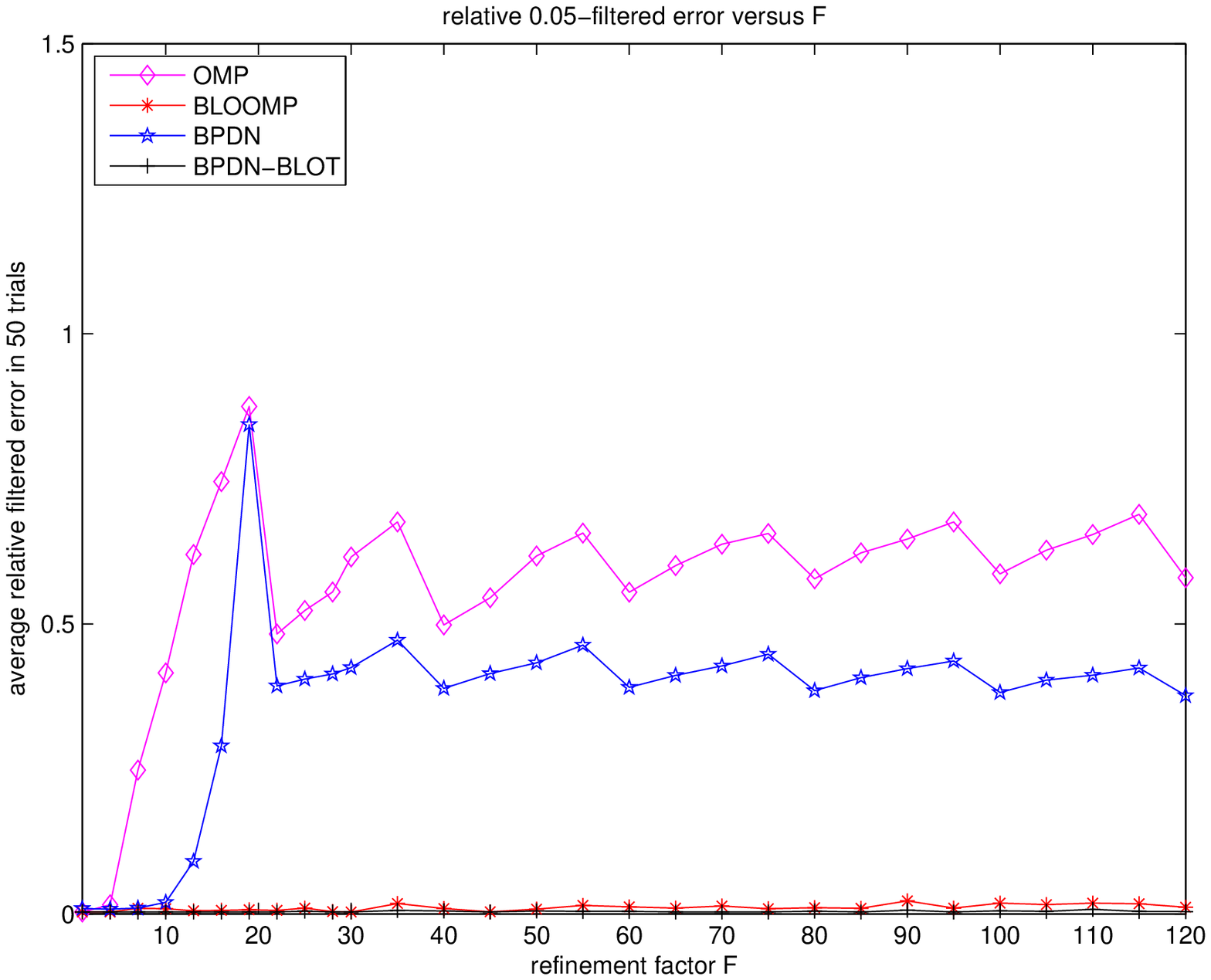}}
      %%%%%%%%%%%%%%%%%%%%%%%%%%%%%%%%%
          \subfigure[SNR=20, $\eta=0$]{
             \includegraphics[width = 1.4in]{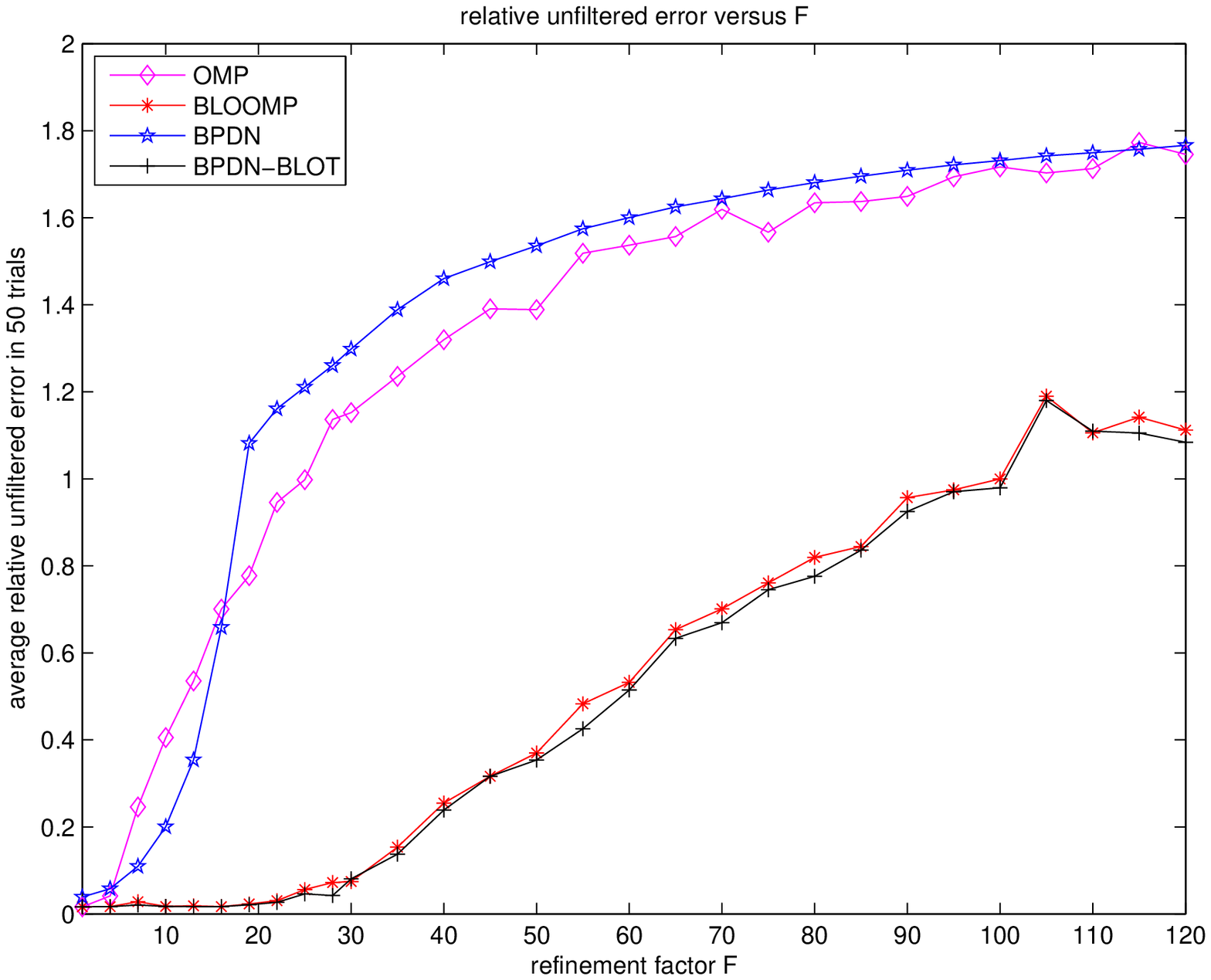}}
        %%%%%%%%%%%%%%%%%%%%%%%%%%%%%%%%%
                  \subfigure[SNR=20, $\eta=0.05\ell$]{
             \includegraphics[width = 1.4in]{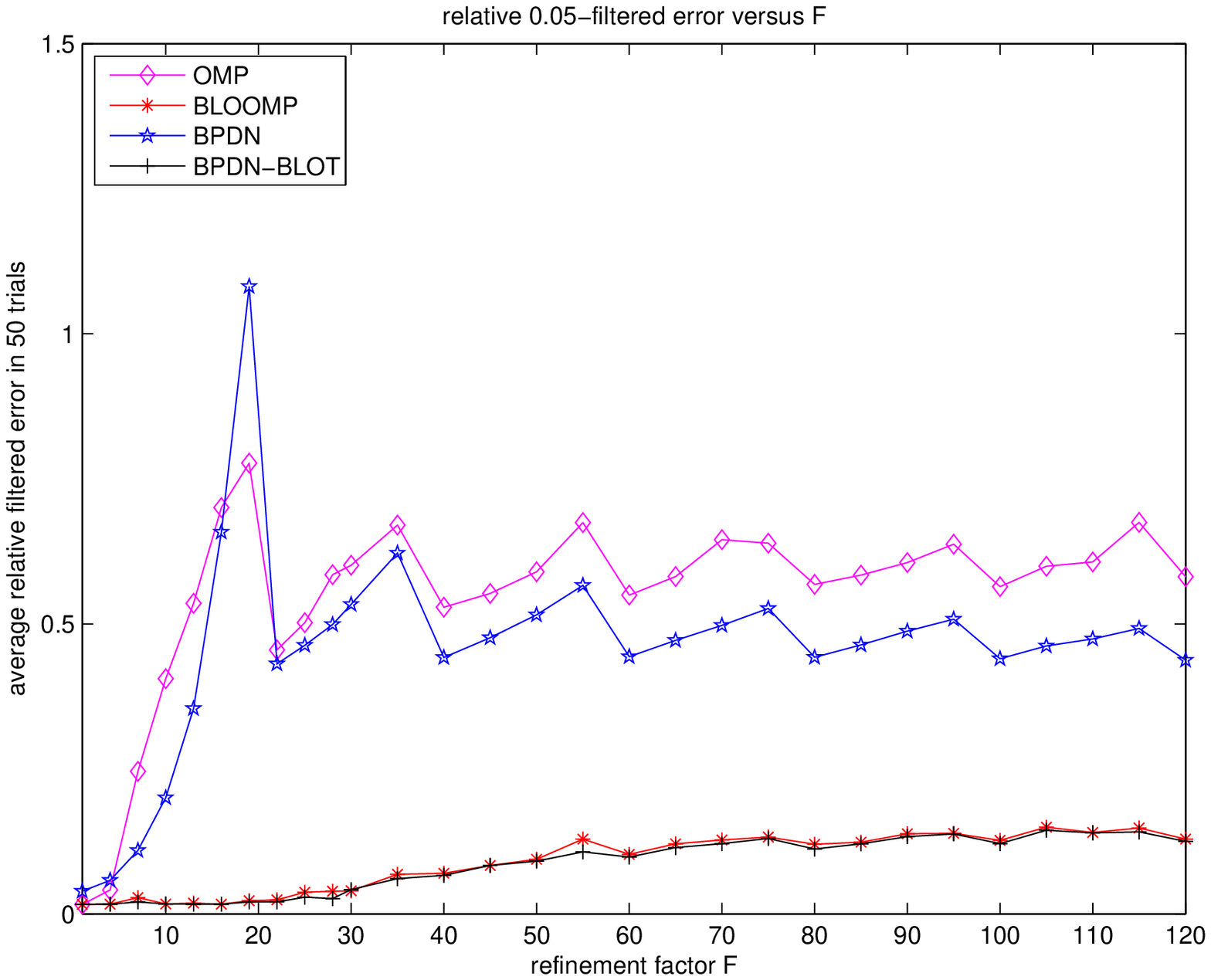}}
        %%%%%%%%%%%%%%%%%%%%%%%%%%%%%%%%%       
           \subfigure[SNR=10, $\eta=0$]{
             \includegraphics[width = 1.4in]{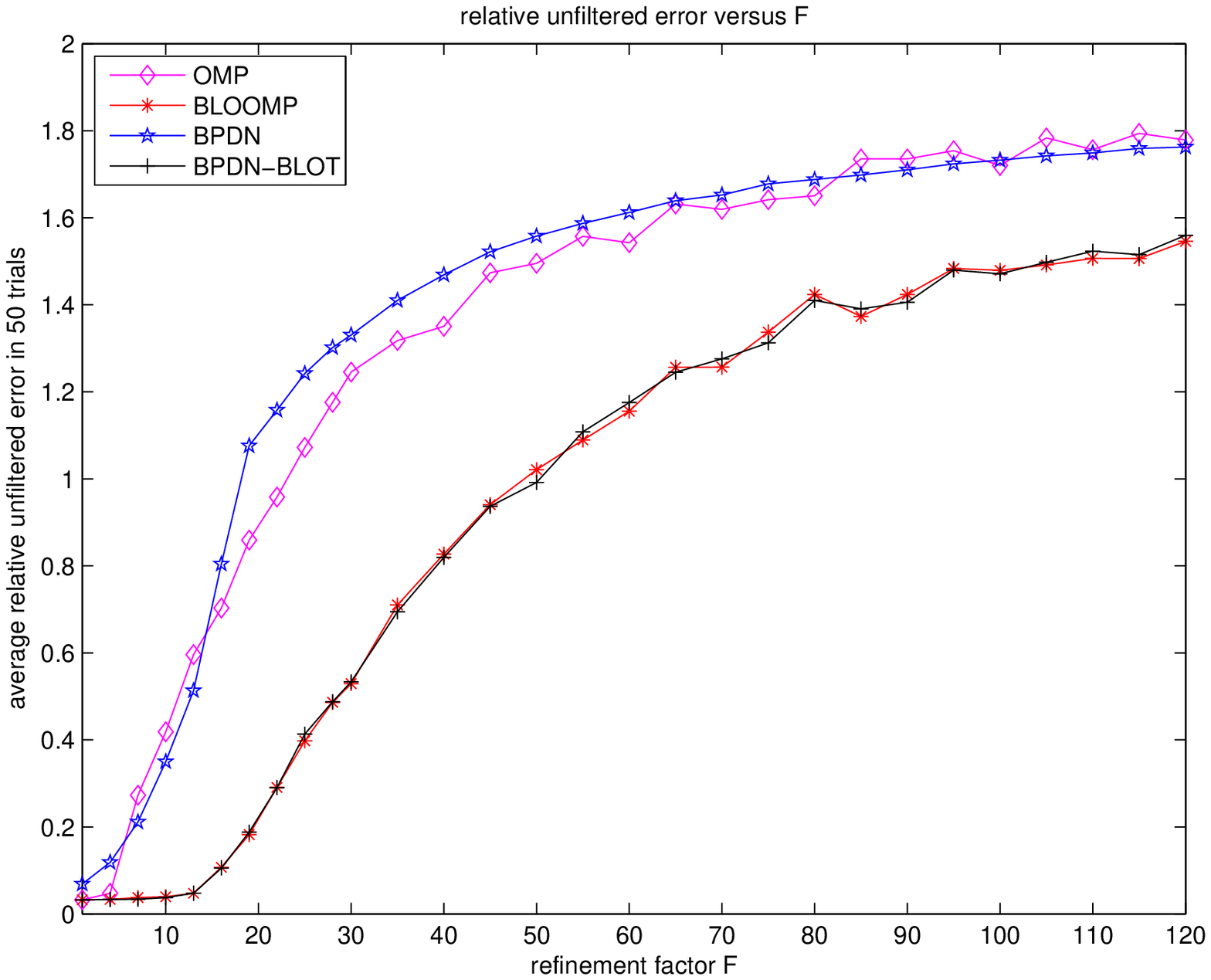}}
             %%%%%%%%%%%%%%%%%%%%%%%%%%%%%%%%%
        \subfigure[SNR=10, $\eta=0.05\ell$]{
             \includegraphics[width =1.4in]{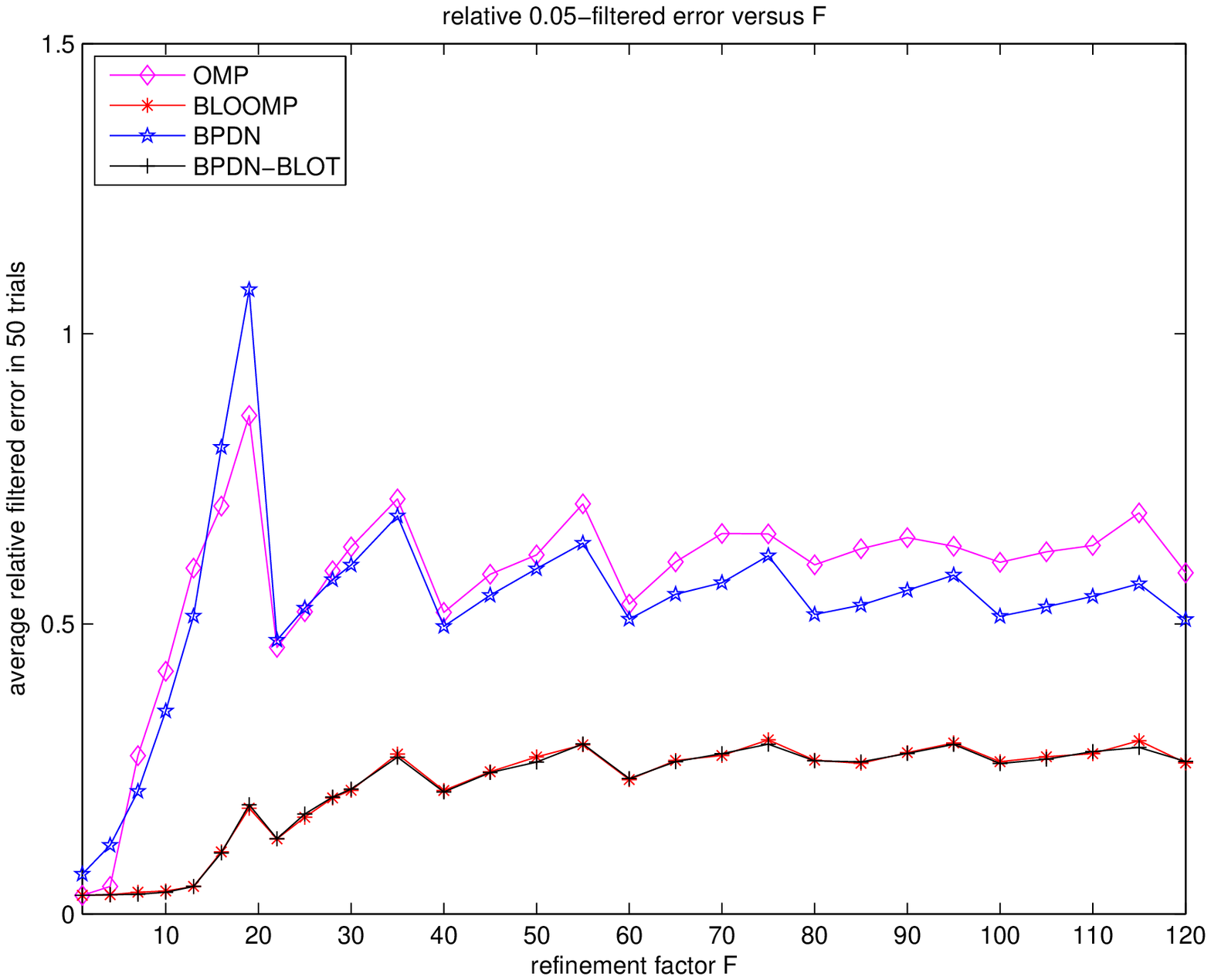}}
              %%%%%%%%%%%%%%%%%%%%%%%%%%%%%%%%%
              \caption{Relative errors in reconstruction by OMP, BLOOMP, BP and BP-BLOT versus the super-resolution factor.
              }
        \label{fig3}
         %% label for entire figure
\end{figure}

 Fig.\ref{fig3} shows the filtered error with $\eta= 0 \& 0.05\ell$ versus $F$ for  $20$ well separated spikes with $1, 5, 10\%$ Gaussian noises. It is noteworthy that the error curves
for OMP and BP are essentially independent of SNR when $F\geq 15$. This may be due to the sensitivity of the algorithms to the round-off error for large $F$.

Also, the power-law amplification (PLA) regimes for OMP and BP are not affected by 
the filtration with  $\eta=0.05\ell$.
 The PLA regime for BP is about $F< 20$ ($F<5$ for OMP)  while the PLA regimes 
 for BLOOMP and BP-BLOT are much milder and slower growing.  
% For SNR=20, the PLA  regimes  for  for BLOOMP and BP-BLOT extend to about $F <100$. 
 
 If we set  the relative error equal to, say thrice  the noise level  as the threshold of successful recovery, then
 in terms of either the unfiltered or filtered error OMP and BPDN
 fail for  $F>10$
while BLOOMP and BP-BLOT succeed for all $F$ in terms of the filtered error, achieving
grid-independent recovery. This remains true  for a lower error threshold 
if the filtration parameter $\eta$ is increased.  
       
        \begin{figure}[t]
  \centering
          \subfigure[OMP]{
            \includegraphics[width = 1.4in]{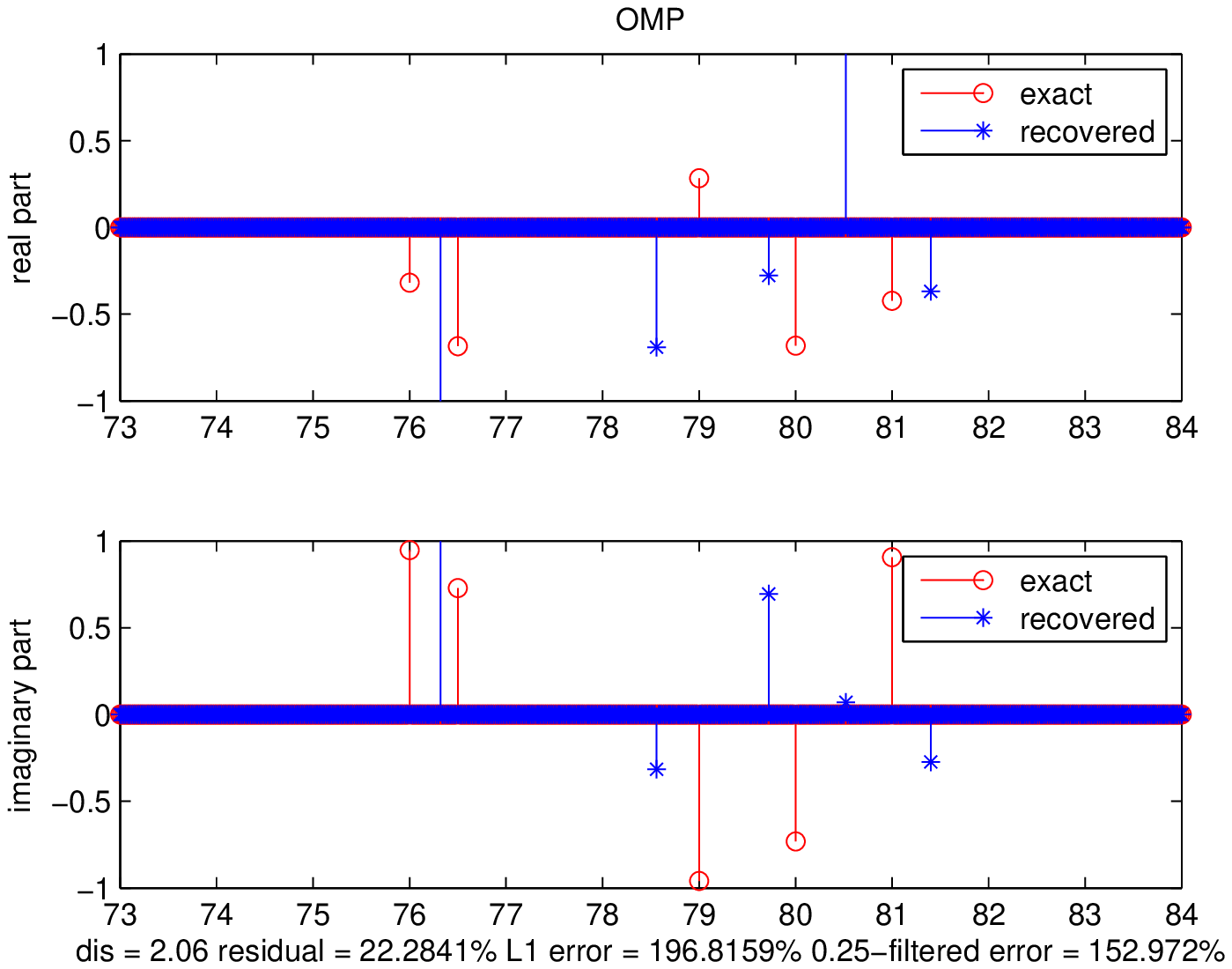}}
        %%%%%%%%%%%%%%%%%%%%%%%%%%%%%%%%%
          \subfigure[BLOOMP]{
            \includegraphics[width = 1.4in]{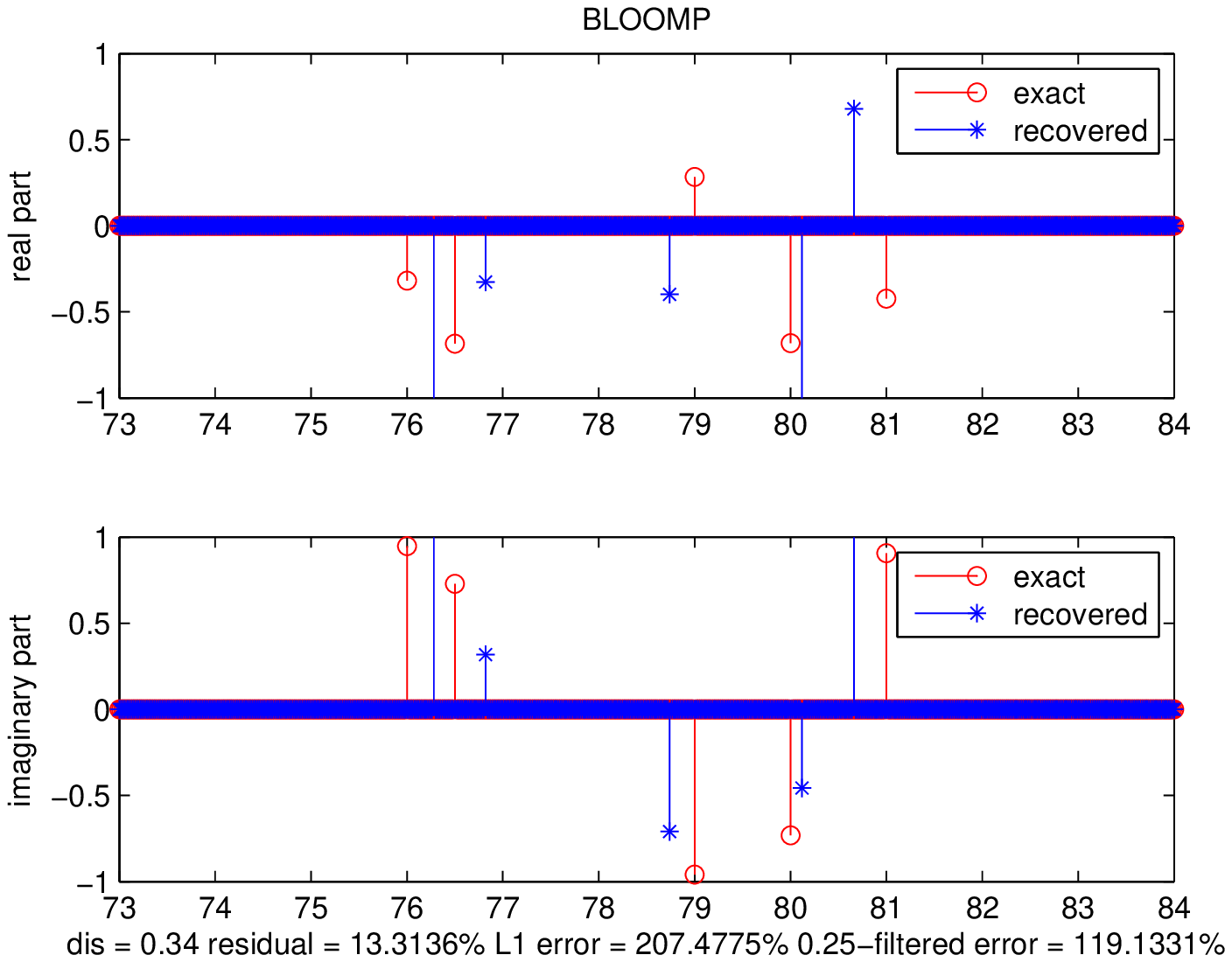}}
      %%%%%%%%%%%%%%%%%%%%%%%%%%%%%%%%%
          \subfigure[BP]{
            \includegraphics[width = 1.4in]{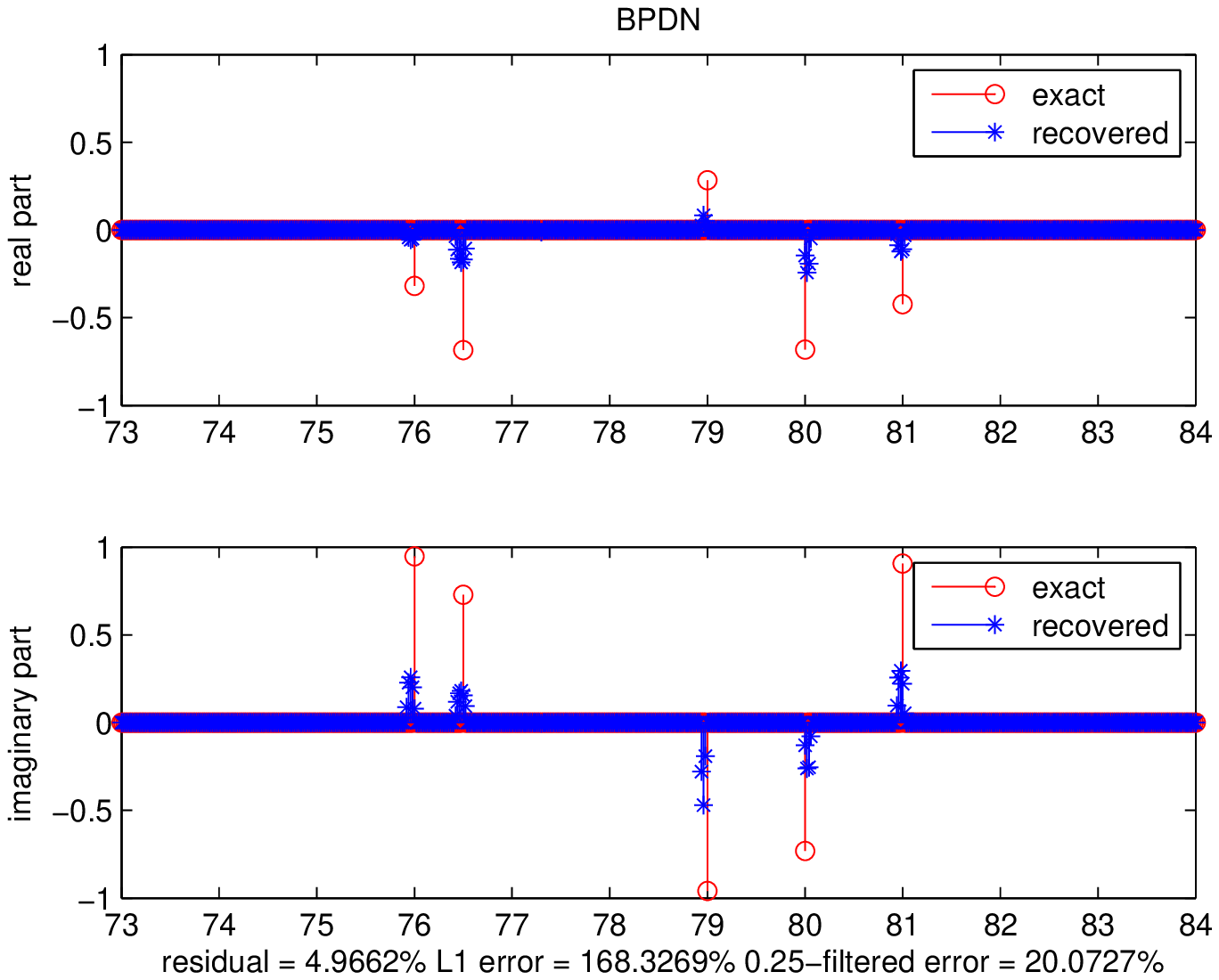}}
        %%%%%%%%%%%%%%%%%%%%%%%%%%%%%%%%% 
         \subfigure[BP-BLOT]{
            \includegraphics[width = 1.4in]{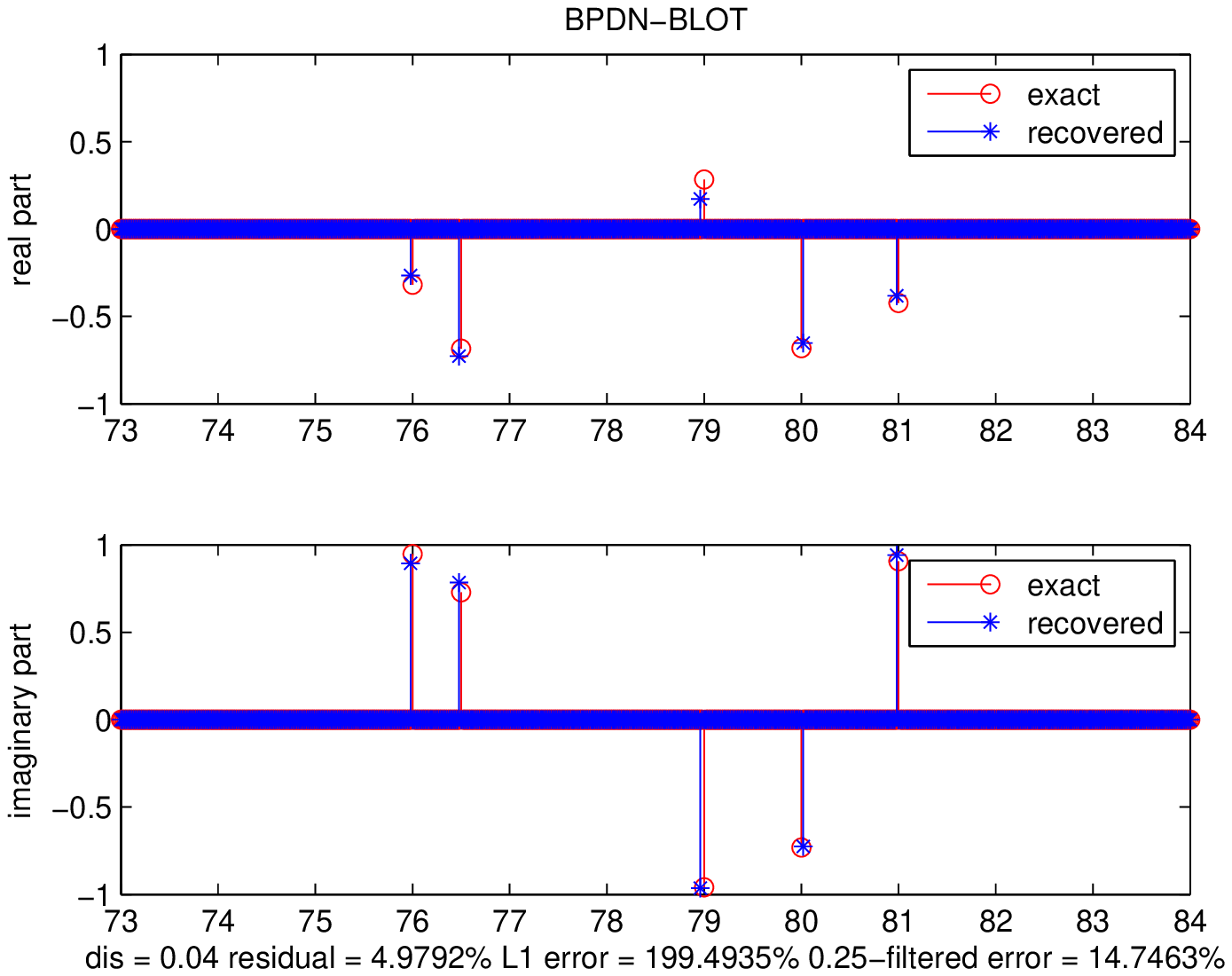}}
         \caption{ Reconstructions  of 5 randomly phased  spikes located at  $76, 76.5, 79, 80, 81\ell$ ($R=5$) with $F=50, \hbox{\rm SNR}=20$.           }
        \label{fig:R5}
        \end{figure}

Next we test the strong form of super-resolution with spikes separated by sub-Rayleigh length.  As mentioned before,
in this case the spikes can fall into one another's coherence band, thus confusing the
recovery.  In this case the excluded zones 
in Algorithms 1, 2 and 3 are not coherence bands 
of previously detected spikes but smaller zones whose size is set to be half the least separation of spikes (Remark 1). Again we set $F=50$
and SNR=20. 

\begin{figure}[t]
  \centering
          \subfigure[OMP]{
            \includegraphics[width = 1.4in]{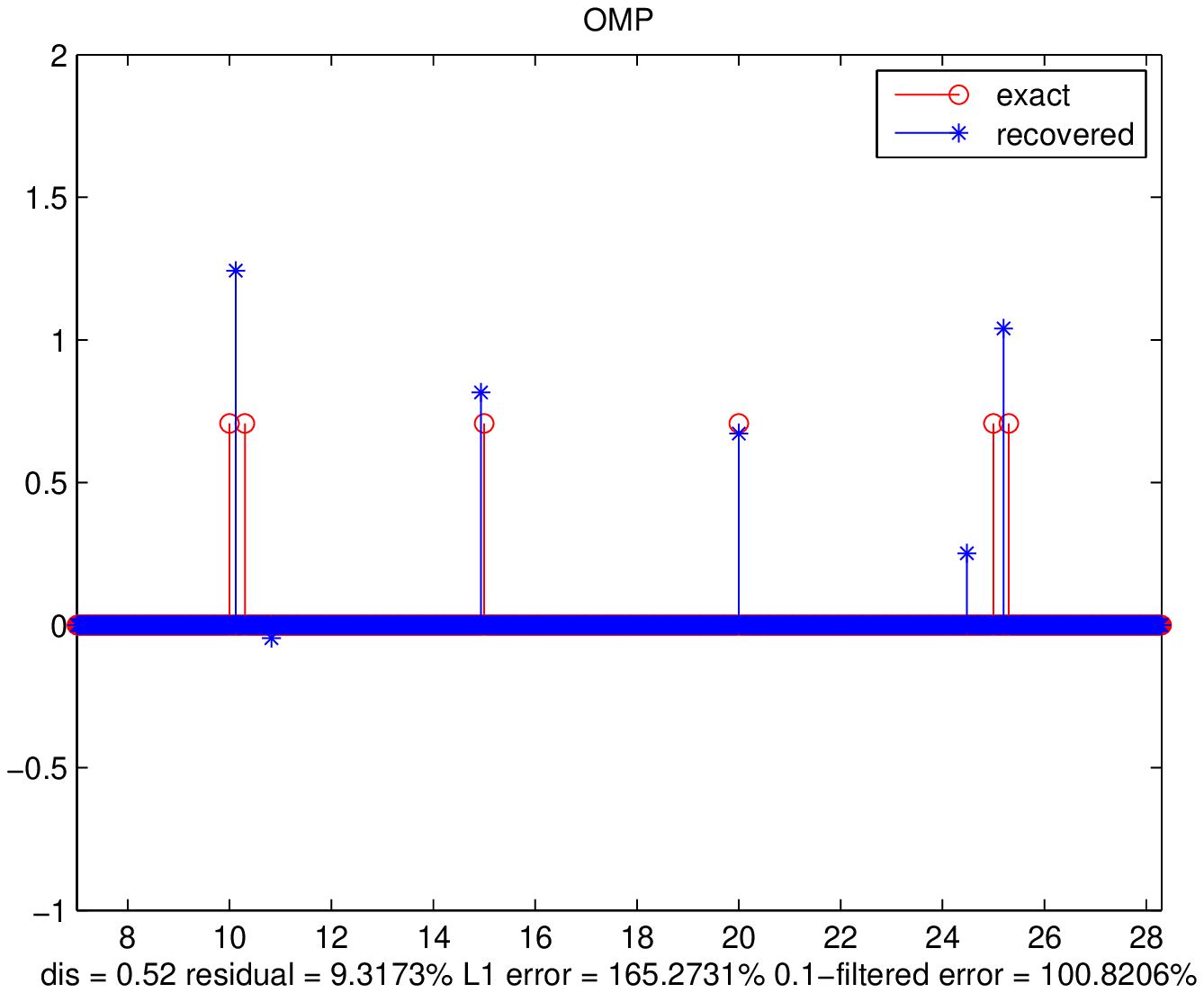}}
        %%%%%%%%%%%%%%%%%%%%%%%%%%%%%%%%%
          \subfigure[BLOOMP]{
            \includegraphics[width = 1.4in]{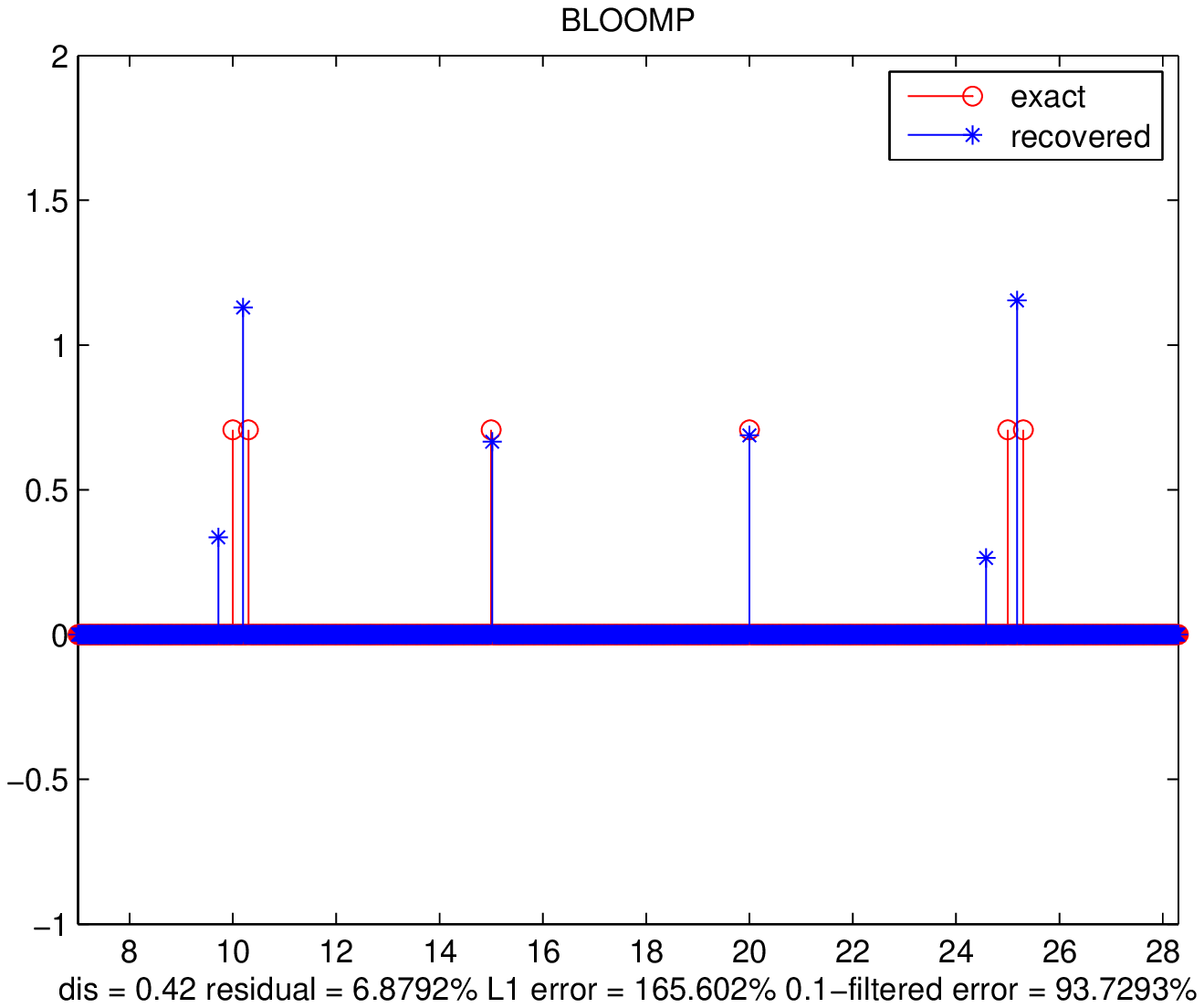}}
      %%%%%%%%%%%%%%%%%%%%%%%%%%%%%%%%%
          \subfigure[BP]{
            \includegraphics[width = 1.4in]{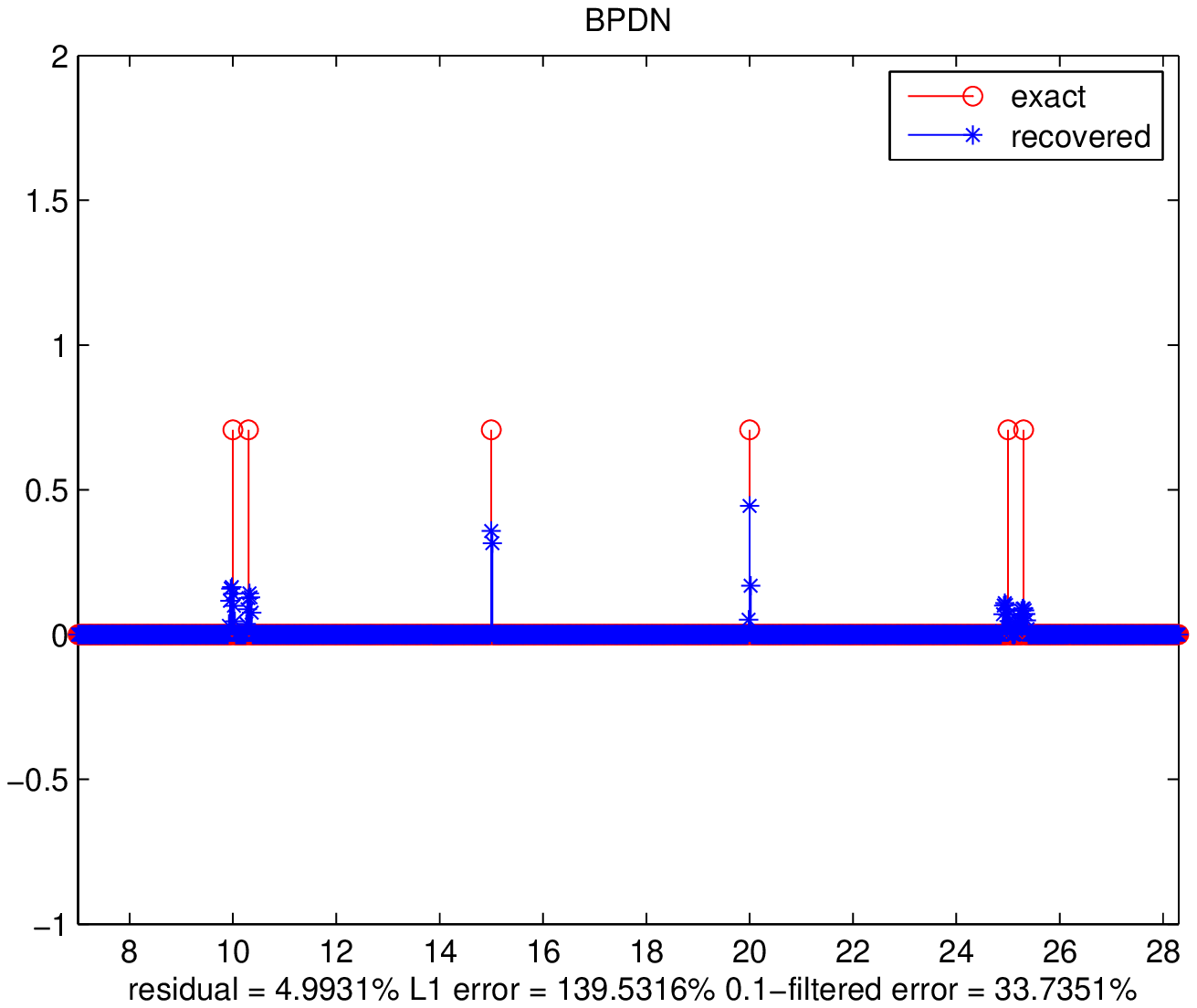}}
        %%%%%%%%%%%%%%%%%%%%%%%%%%%%%%%%% 
         \subfigure[BP-BLOT]{
            \includegraphics[width = 1.4in]{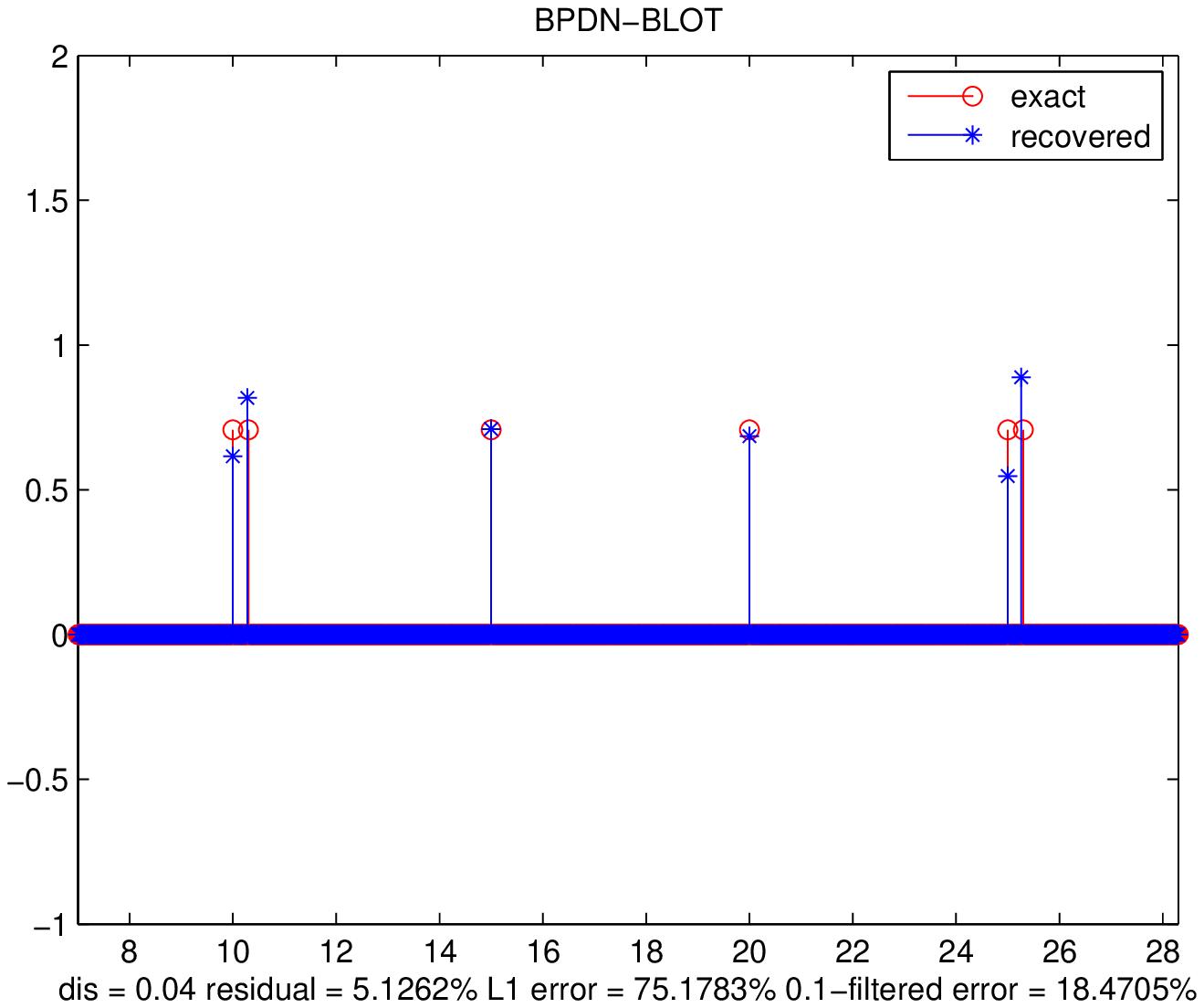}}
         \caption{ Reconstructions of the real parts of 5 complex spikes located at  $10, 10.3, 15, 20, 25, 25.3\ell$ ($R= 6$) with $F=50, \hbox{\rm SNR}=20$.          }
        \label{fig:R2}
        \end{figure}

Fig.\ref{fig:R5} shows the reconstructions of 5 randomly phased spikes ($S=\{76, 76.5, 79, 80, 81 \ell\}$) while
Fig.\ref{fig:R2} shows the reconstructions of 6 complex spikes of positive real parts ($S=\{10, 10.3, 15, 20, 25, 25.3\ell\}$) 
by OMP, BP, BLOOMP and BP-BLOT.  According to the definition (\ref{ray}) the former set has
Rayleigh index 5 and the latter set has Rayleigh index 6. Clearly, only 
BP-BLOT performs reasonably well in both cases. 

Several observations are in order. First, when some spikes are separated by  less
than $\ell$, the BLO technique may not improve on the  OMP reconstructions.
%in terms of  any of the three performance metrics: the Bottleneck distance, the unfiltered and
%filtered errors. 
Second, the BLOT technique improves on the BP reconstructions, especially for the Bottleneck distance of support offset, achieving the accuracy of
$0.04\ell$ in both Fig.\ref{fig:R5} and Fig.\ref{fig:R2}. 
Third,  the filtered errors  for BP-BLOT (15\% with $\eta=0.25\ell$ in Fig.\ref{fig:R5}(d) and
$18\%$ with $\eta=0.1\ell$ in Fig.\ref{fig:R2}(d)) are much smaller
than the unfiltered errors (199\% in Fig.\ref{fig:R5}(d) and 75\% in Fig.\ref{fig:R2}(d)).  Fig.\ref{fig:R5} (d) most clearly 
demonstrates  the inadequacy of   the unfiltered norm as the error metric for spike recovery. 

Zooming in on the two pairs of closely spaced spikes in Fig.\ref{fig:R2}(c) will
give us a better sense of how the BLOT technique achieves the feat of super-resolution (see Fig.\ref{fig:zoom}). 
BLOT capitalizes on the tendency of  BP spikes to mushroom around
the true spikes and use the extra prior information (sparsity and minimum separation)
to prune and grow the reconstruction. 

Finally, unlike the case of Rayleigh index 1 in Fig.\ref{fig:R1}, the BLOT  technique slightly increases  the residuals (from 4.97\% to 4.98\% in Fig.\ref{fig:R5} and 4.99\% to 5.13\% in Fig.\ref{fig:R2}) to achieve the super-resolution effect.  In contrast, the BLO technique always reduces the residuals which may not help recovery of spikes separated by sub-Rayleigh length. \\

 \commentout{
\begin{figure}[t]
  \centering
          \subfigure[OMP]{
            \includegraphics[width = 1.5in]{superresolution/OMPSpike6F100Noise5.eps}}
        %%%%%%%%%%%%%%%%%%%%%%%%%%%%%%%%%
          \subfigure[BLOOMP]{
            \includegraphics[width = 1.5in]{superresolution/BLOOMPSpike6F100Noise5.eps}}
      %%%%%%%%%%%%%%%%%%%%%%%%%%%%%%%%%
          \subfigure[BP]{
            \includegraphics[width = 1.5in]{superresolution/BPDNSpike6F100Noise5.eps}}
        %%%%%%%%%%%%%%%%%%%%%%%%%%%%%%%%% 
         \subfigure[BP-BLOT]{
            \includegraphics[width = 1.5in]{superresolution/BPDNBLOTSpike6F100Noise5.eps}}
         \caption{ Reconstructions  of spikes with $R= 3$ (minimum distance $0.2 \ell $), $F=100, \hbox{\rm SNR}=20$.
          }
        \label{fig:R3}
         %% label for entire figure
\end{figure}
}

\centerline{\footnotesize V. CONCLUSION}
We have argued  that the discrete unfiltered norm is not
a proper error metric for spike recovery since the offset of the support
recovery is not accounted for and that a filtered error norm
may be used instead. 

We have demonstrated that 
both BLOOMP and BP-BLOT can recover
well separated spikes in terms of the filtered error norm
as well as localize the spikes within a few percent of
the Rayleigh length, {\em independent of the super-resolution factor}. This is a weak form of super-resolution.

When some spikes are closely spaced below the Rayleigh length, only BP-BLOT 
can  localize the spikes within a few percent of $\ell$.  
%although  the  filtered or unfiltered errors remain quite substantial. 
%The BLOT technique improves
%only slightly the quality of amplitude recovery, resulting  in
%small reduction of both the  filtered and unfiltered errors. 
The performance  can be further enhanced by increasing
the number of Fourier data within the same bandwidth,  moving from the under-sampling 
 to the full and even  over-sampling regimes (not shown).

Our numerical tests show that the super-resolution factor
and the Rayleigh index are not the only factors
at play.  The minimum separation,  the range of spike values and
overall configuration also affect super-resolution results. 
     \begin{figure}[t!]
  \centering          \subfigure[spikes at 10, 10.3$\ell$]{
            \includegraphics[width = 1.4in]{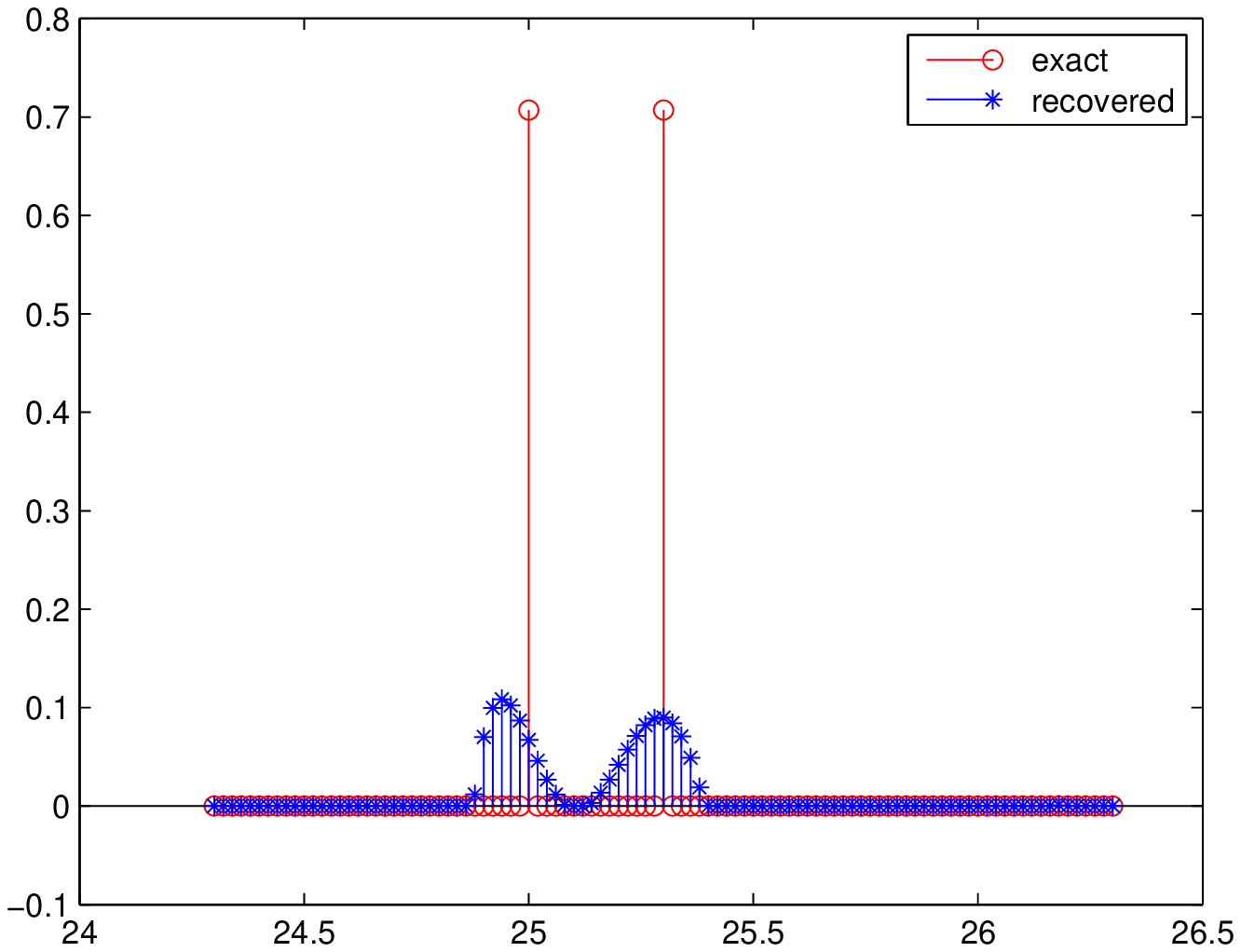}}
        %%%%%%%%%%%%%%%%%%%%%%%%%%%%%%%%%
          \subfigure[spikes at 25, 25.3$\ell$]{
            \includegraphics[width = 1.4in]{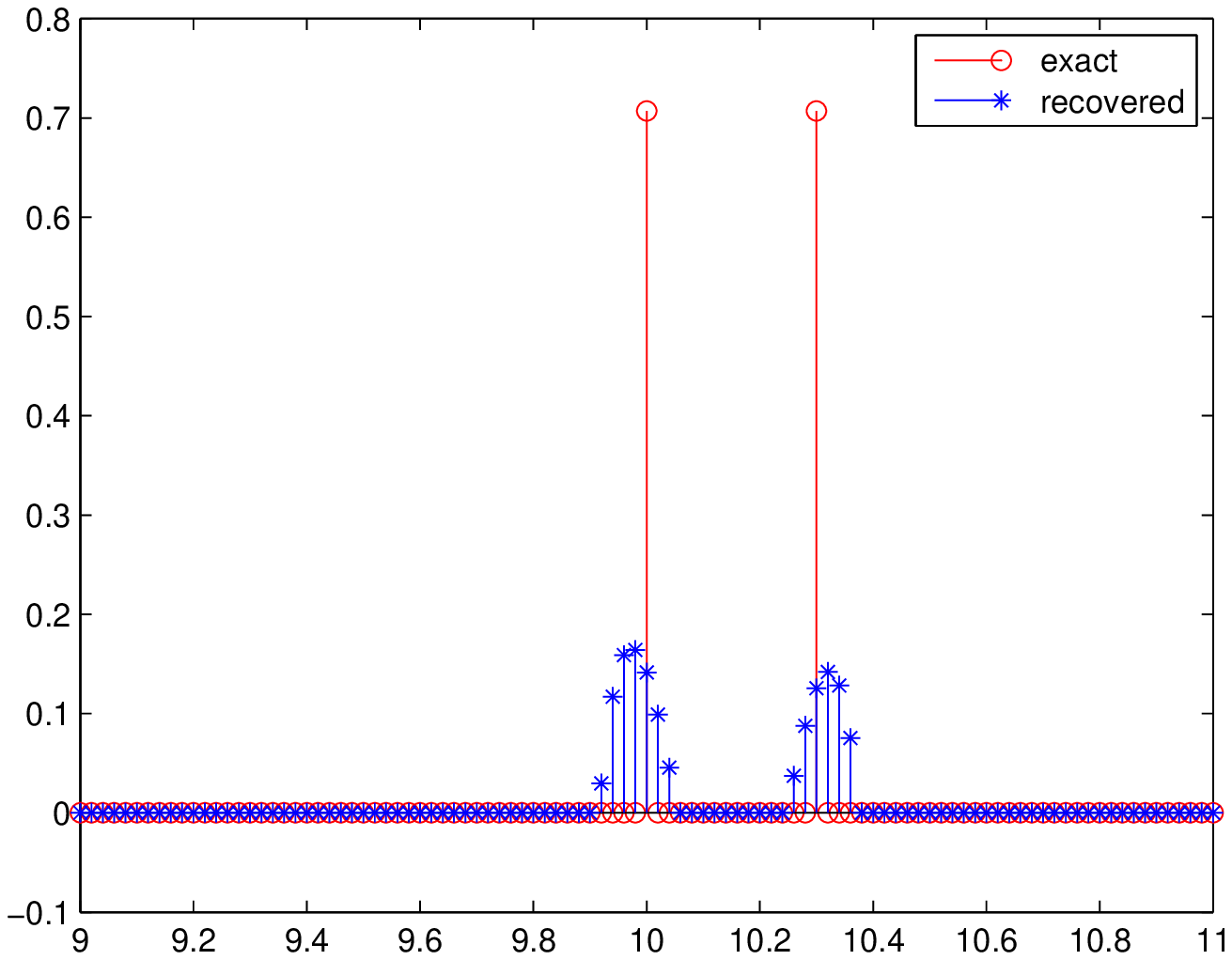}}
\commentout{
          \subfigure[spikes at 10, 10.3]{
            \includegraphics[width = 1.4in]{BushTwoSpike1.eps}}
        %%%%%%%%%%%%%%%%%%%%%%%%%%%%%%%%%
          \subfigure[spikes at 25, 25.3]{
            \includegraphics[width = 1.4in]{BushTwoSpike2.eps}}
      %%%%%%%%%%%%%%%%%%%%%%%%%%%%%%%%
      }
         \caption{Zoom-ins of the closely spaced spikes in Fig.\ref{fig:R2}(c).  }
        \label{fig:zoom}
        \end{figure}

\end{document}